\documentclass[10pt]{iopart}
\usepackage{iopams}
\usepackage{pfnote}
\usepackage{graphicx}
\usepackage{amssymb}
\usepackage{bbm}
\usepackage[utf8]{inputenc}
\usepackage[normalem]{ulem} 

\usepackage{xspace} 
\usepackage{color}
\usepackage{footnote}

\usepackage{hyperref}

\usepackage{braket}
\usepackage{bm}

\newcommand{\Graz}{Institute of Theoretical and Computational Physics, Graz University of Technology, 8010 Graz, Austria}
\newcommand{\Rom}{Dipartimento di Fisica, Sapienza Universit\`{a} di Roma, P.le A. Moro, 2 - 00185 ROMA, Italy}

\newcommand{\opa}{{\mathop{\hat{a}}}}
\newcommand{\opad}[1]{{{\mathop{\hat{a}_{#1}^\dagger}}}} %
\newcommand{\opn}{{\mathop{\hat{n}}}}
\newcommand{\opH}{{\mathop{\hat{H}}}}

\newcommand{\opV}{{\mathop{\hat{V}}}}

\newcommand{\rumcomm}[1]{{\color{black} #1}}

\newcommand{\quot}[1]{ "{#1}"}

\begin{document}

\title{First-principles molecular transport calculation for the benzenedithiolate molecule}

\author{M.~Rumetshofer$^1$, G.~Dorn$^1$, L.~Boeri$^{1,2}$, E.~Arrigoni$^1$, W.~von der Linden$^1$}

\address{$^1$\Graz}
\address{$^2$\Rom}

\ead{m.rumetshofer@tugraz.at}

\date{\today}

\begin{abstract}
A first-principles approach based on Density Functional Theory and Non-Equilibrium Green's functions is used to study the molecular transport system consisting of benzenedithiolate connected with monoatomic gold and platinum electrodes.
Using symmetry arguments we explain why the conductance mechanism is different for gold and platinum electrodes.
We present the charge stability diagram for the benzenedithiolate connected with monoatomic platinum electrodes including many-body effects in terms of an extended Hubbard Hamiltonian
and discuss how the electrodes and the many-body effects influence the transport properties of the system.
\end{abstract}

\section{Introduction}

First-principles calculations in the field of molecular electronics represent a major challenge in computational physics.
{In order to build electronic devices based on organic molecules it is crucial to understand the underlying charge transport phenomena in detail to be able to simulate the actual behaviour \cite{Joachim_MolElRatner_2005, Cuevas_MolecularElectronics_2010}. However, }
it is a challenging task
to predict quantitatively the experimental characteristics of molecular junctions.
First of all, in many cases the charge transport properties of a system depend on the detailed contact geometry \cite{Pontes_AuBDTAu_2011, Cuniberti_Contacts_2002, Kim_IETS_2011}, which is experimentally difficult to identify.
But even if all atomic positions were known, non-equilibrium calculations are always based on approximations which affect the results.
Different theoretical methods exist for different transport regimes.
In the \textit{coherent} regime electrons proceed elastically, without exchanging energy.
This is the case if there is a strong coupling between the leads and the molecule, i.e. if
\begin{equation}
 \Gamma \gg \Delta E, ~U,~ k_BT ,
\end{equation}
here, $\Gamma$ is the coupling between the leads and the molecule, $\Delta E$ the level spacing, $U$ the Hubbard interaction parameter and $k_BT$ the temperature. 
In the coherent regime, the energy levels of the electrodes and the molecule are hybridized and charge quantization is suppressed
and  the Landauer formalism becomes exact.
The combination of Density Functional Theory and non-equilibrium Green's functions (DFT+NEGF) has become a standard tool for coherent ab-initio transport calculations \rumcomm{but there are also other approaches like the Lippmann-Schwinger method \cite{Ventra_BDT_2000} or the wave-function matching technique \cite{Khomyakov_FDM_2004}}.
In the \textit{incoherent} regime the time that an electron spends on the system is sufficiently long for it to interact with other particles (electrons, phonons, etc.).
{In the case of electron-electron interaction we have}
\begin{equation}
\Gamma < \Delta E,~ U 
\end{equation}
{for the incoherent regime.}
If the leads are weakly coupled to the system then there is charge quantization and the electrons propagate by sequential tunnelling. 
This incoherent regime can be modelled by a Hubbard-Hamiltonian for describing the molecule and non-interacting tight-binding leads.
Non-equilibrium cluster perturbation theory (n-CPT) \cite{ba.po.11, nu.he.12} or the variational cluster approach (VCA) \cite{kn.li.11, nu.he.12} are techniques for solving this model accurately in the coherent limit.
In the case of weak coupling between the molecule and the leads master equation (ME) \cite{Breuer_ME_2007} approaches became useful.
The auxiliary master equation approach (AMEA) \cite{Arrigoni_AMEA_2013, Dorda_AMEA_2014, do.ga.15} is in principle exact, also for strong coupling between the molecule and the leads, but is limited to systems with only a view interacting sites in the central region.
In order to calculate transport properties from first principles one can combine these techniques with DFT
and carry out a charge self-consistency procedure in non-equilibrium, which is often neglected for simplicity.
In our work we apply DFT+NEGF and an extended version of DFT+NEGF, which we denote by DFT+CPT, for calculations including electron-electron interactions.
The approximation in DFT+CPT is to approximate the self-energy of the system by the self-energy of the isolated central region.
The method allows treating larger molecules in the central region and is accurate in the coherent limit.
A more sophisticated but numerically challenging technique for calculating transport properties is the combination of DFT with GW \cite{Hedin_GW_1965, Thygesen_BDT_2011, Thygesen_GW_2008}. 
DFT can also be used in combination with the dynamical mean field theory (DMFT) \cite{Jacob_DMFT_2015}, 
with ME \cite{Ryndyk_BDT_2013} if the system is in the weak coupling regime or 
even with ME+CPT \cite{Nuss_MECPT_2015}.

Although the 1,4-benzenedithiolate (BDT) molecule is usually considered in the field of molecular electronics there are still severe discrepancies between experimental \cite{Reed_BDTExperiment_1997, Loertscher_BDTExperiment_2007, Song_exp_2009, Xiao_exp_2004, Tsuitsui_exp_2009, Kiguchi_exp_2010} and theoretical results \cite{Xie_ExpTheoOligopheneDithiol_2015, Ryndyk_BDT_2013, Kim_BDTTrans_2011, Kondo_BDTchannel_2006, Xue_bandlineup_2001,Ventra_BDT_2000}.
The calculated conductances tend to be much higher than the measured ones.
{Proposed reasons are the sensitivity of the conductance to the detailed geometry of the junction \cite{Kondo_BDTchannel_2006, Ventra_BDT_2000}, 
hydration of the BDT molecule to benzenedithiol \cite{Souza_thiolthiolate_2014, Thygesen_BDT_2011} and correlation effects \cite{Pontes_AuBDTAu_2011, Sanvito_SIC_2007}.}

\rumcomm{The purpose of this work is to to study the transport properties of the BDT molecule connected to monoatomic metal chains as leads.
This system allows to study the impact of the junction geometry on  the conductance
and how transport is affected by low dimensional leads.
Many publications \cite{Kim_BDTTrans_2011, Xie_ExpTheoOligopheneDithiol_2015, Xue_bandlineup_2001} explain the transport through the gold-BDT-gold system with a single-level model where the conductance is carried by a single transport channel corresponding to the HOMO orbital of the BDT molecule.
Ryndyk et al. \cite{Ryndyk_BDT_2013} performed DFT+ME calculations including strong electron correlations and got a multi-scale Coulomb blockade.
We will also  address the impact of strong correlations. Our results obtained for the single level
model for the BDT molecule will be compared with the results
for the model studied by Ryndyk et al. \cite{Ryndyk_BDT_2013}.}

The paper is organized as follows.
\Sref{Methods} introduces briefly the methodology and is divided in two parts.
The first part describes how the parameters in the extended Hubbard Hamiltonian are obtained from first-principles,
while the second part presents a technique based on NEGF and CPT for solving this model.
In \sref{Applications} the method is applied to a gold-BDT-gold and a platinum-BDT-platinum system.
The conclusions are given in \sref{Conclusions}.

\section{Methods}
\label{Methods}

\subsection{First-principles calculation of the model parameters}

\begin{figure}[t]
\includegraphics[width=1.0\columnwidth,angle=0]{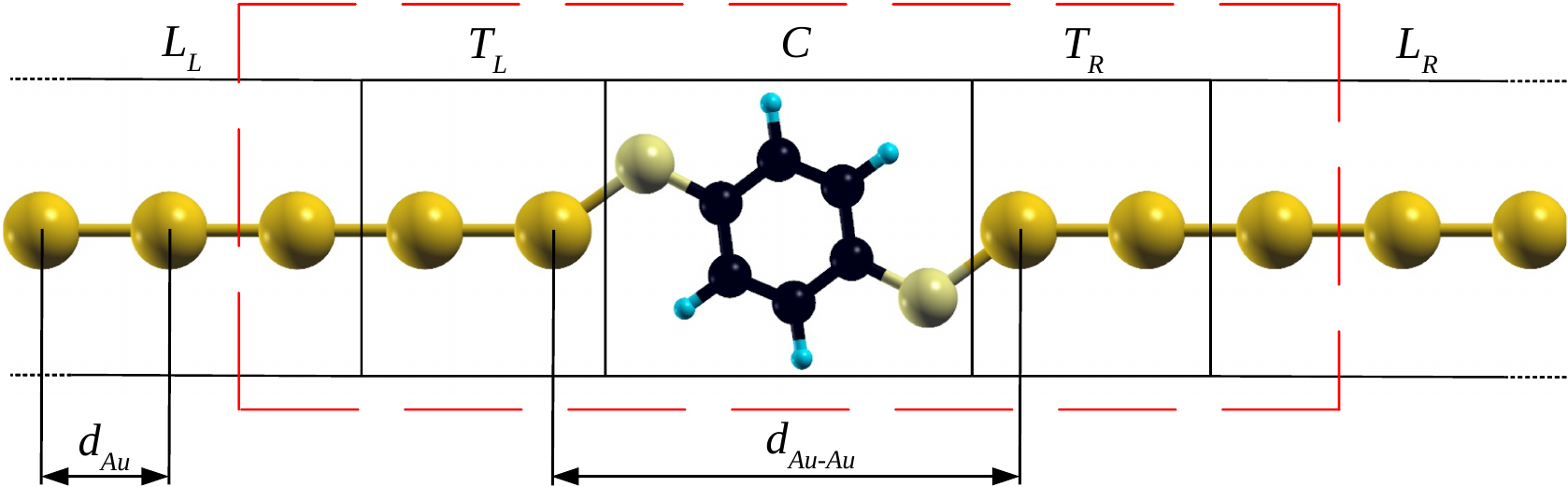}
\caption{Molecular system consisting of a left and a right lead ($L_L$ and $L_R$), transition layers ($T_L$ and $T_R$) and the central region ($C$). The picture is drawn with XCrySDen \cite{Kokalj_xcrysden_2003}.}
\label{fig:trans}
\end{figure}

The system we are considering consists of a central region (C) attached to left and right leads, see \fref{fig:trans}.
To determine the tight-binding (TB) parameters of the system we employ the pseudopotential plane-wave code {\sc Quantum Espresso} \cite{QE_2009}.
The unit cell, red dashed line in \fref{fig:trans}, consists of the central region, left and right  transition layers ($T_L$, $T_R$), also called buffer regions, and the surface part of the remaining leads ($L_L$,$L_R$).
The size of the transition layers is chosen such that the electronic properties of the outermost atoms in the unit cell do not change anymore when the unit cell
is increased. 
This ensures  that the central region has negligible  impact on the periodicity of the remaining leads \cite{Calzolari_WannierConductance_2004, Shelley_WannierTrans_2011}.
For all DFT calculations in this paper we have used the Perdew Zunger (PZ) exchange correlation functional within the local density approximation (LDA) and non-relativistic ultrasoft pseudopotentials from the Standard Solid-State Pseudopotentials (SSSP) library (PSlibrary 0.3.1) \cite{DalCorso_PseudoPotential_2014}.

We have used Maximally Localized Wannier Functions (MLWF) \cite{Marzari_Wannier_1997, Souza_Wannier_2001} to get localized orbitals and to set up the TB Hamiltonian \cite{Calzolari_WannierConductance_2004, Shelley_WannierTrans_2011}.
All Wannier transformations are done with {\sc wannier90} \cite{Mostofi_wannier90__2014} \rumcomm{which produces orthogonal MLWF. Therefore the derivation of the NEGF formalism in section \ref{NEG} is demonstrated only for orthogonal bases. A derivation for non-orthogonal bases in the coherent regime can be found in \cite{Krstic_nonortho_2002}.} 
In the localized basis the Hamilton operator of the unit cell is then given by
\begin{equation}
\opH = \left( \begin{array}{ccccc}
\opH_{L_L} & \opV_{L_LT_L} & & & \opV_{L_LL_R} \\
 \opV_{L_LT_L}^\dagger  & \opH_{T_L} & \opV_{T_LC} &  \\
  & \opV_{T_LC}^\dagger & \opH_C & \opV_{CT_R} &  \\
  &  &  \opV_{CT_R}^\dagger & \opH_{T_R} & \opV_{T_RL_R} \\
  \opV_{L_LL_R}^\dagger &  & & \opV_{T_RL_R}^\dagger & \opH_{L_R}  \\
\end{array} \right) \;.
\label{eq_hameLCR}
\end{equation}
The outermost atoms of the unit cell are periodically continued and form the non-interacting 
leads ($L_L$, $L_R$) in the TB model. 
In this way we determine all single-particle parameters of the Hamiltonian.

We describe the electron-electron interaction by an extended Hubbard Hamiltonian in the central region, which has the form
\begin{equation}
 \opH_C = \opH_0 + \opH_1
\end{equation}
where $\opH_0$ is the single-particle part and $\opH_1$ the interaction part of the model Hamiltonian.
We neglect correlation effects in the leads ($L_L$,$T_L$,$L_R$,$T_R$) 
and we expect correlation effects to be important only in the central region, because on the molecule the wave functions are more localized.
The single-particle part of the model Hamiltonian in the central region in second quantization is given by
\begin{equation}
\opH_0 = \sum_{ij}^N \tilde t_{ij} \opad{i}\opa_j 
\label{eq_hamiltonian_0}\;.
\end{equation}
The indices $i=\{ \bar{i},\sigma \}$ and $j=\{ \bar{j},\sigma' \}$ include the wave function indices $\bar{i}$ and $\bar{j}$ and the spins $\sigma$ and $\sigma'$.
The operator $\opad{i}$ creates an electron on orbital $\Psi_i({\bf r})$ which we obtain by the MLWF construction. 
The hopping parameters $\tilde{t}_{ij}$ are
\begin{equation}
\tilde  t_{ij} =\{Q E Q^\dagger\}_{\bar{i}{\bar{j}}}\;,
\end{equation}
where $Q$ is the matrix which diagonalizes the TB Hamiltonian and $E$ are the corresponding eigenenergies. 
The interaction part of the model Hamiltonian is given by
\begin{equation}
\opH_1 = \frac{1}{2}\sum_{i\neq j}^N U_{ij} \opn_{i}\opn_{j}\;.
\label{eq_hamiltonian_1}
\end{equation}
The density-density term form of the Hubbard model is only justified if \rumcomm{the non-density-density term is small} 
which also requires that the orbitals are maximally localized. 
The interaction parameters $U_{ij}$ are calculated via
\begin{equation}
U_{ij} = \int d^3r \int d^3r' \left|  \Psi_i({\bf r}) \right| ^2 \left|  \Psi_j({\bf r'}) \right| ^2 U({\bf r}, {\bf r'})
\label{eq_U2}
\end{equation}
with the screened Coulomb potential
\begin{equation}
U({\bf r}, {\bf r'}) = \frac{1}{4\pi\epsilon_0 \eta |{\bf r}-{\bf r'} |}
\end{equation}
where $\eta$ is the screening. According to \cite{Ryndyk_BDT_2013} we set the screening to be constant.
The Random Phase Approximation (RPA) \cite{Bohm_RPA_1953} would be a possibility to determine the screening consistently.

The electron correlations already included in DFT, have to be subtracted in order to avoid double counting.
Here we choose the around mean-field approximation (AMF) \cite{Karolak_DoubleCounting_2010}, because  it generally gives good results for weakly correlated systems.
That is, we have subtracted the Hartree term $\Delta t_{\bar{i}\bar{i}}$ of the interaction part from the bare hopping $\tilde{t}_{ij}$. This approach is justified as follows. If we apply the LDA approximation to the model Hamiltonian we end up with the same band structure as that is obtained from the original band structure calculation. The double counting correction leads
to modified hopping parameters $\tilde  t_{ij}\to t^{C}_{ij}$ with 
\begin{equation}
t_{\bar{i}{\bar{j}}}^C = \tilde{t}_{\bar{i}{\bar{j}}} - \Delta t_{\bar{i}\bar{i}} \delta_{\bar{i}{\bar{j}}} \;.
\end{equation}
The Hartree term is given by
\begin{eqnarray}
\Delta t_{\bar{i}\bar{i}} &\approx U_{\bar{i}\bar{i}}n_{\bar{i}}^0 + \sum_{{\bar{m}}\neq \bar{i}} 2 U_{\bar{i}{\bar{m}}}n_{\bar{m}}^0 \nonumber\\
&= \sum_{\bar{k}}^{N_{occ}} \sum_{\bar{m}} \left| \{ Q^\dagger \} _{{\bar{k}}{\bar{m}}} \right| ^2 \left( U_{\bar{i}\bar{i}}\delta_{\bar{i}{\bar{m}}} + 2U_{\bar{i}{\bar{m}}}(1-\delta_{\bar{i}{\bar{m}}}) \right)
\end{eqnarray}
where $n_{\bar{i}}^0$ is the equilibrium population of the $i$-th orbital obtained in the DFT calculation.
The model Hamiltonian for the central region has finally the form
\begin{equation}
\opH_C = \sum_{ij}^N  t^{C}_{ij} \opad{i}\opa_j 
+\frac{1}{2}\sum_{i\neq j}^N U_{ij} \opn_{i}\opn_{j}\;.
\label{eq_hamiltonian}
\end{equation}
The full self-consistent non-equilibrium transport calculation could be done by
feeding the non-equilibrium potential, obtained from the non-equilibrium densities in the transport calculation, back into DFT
and repeat this procedure until self-consistency is reached \cite{Quian_quasi1delectrodes_2007, Kurth_TDFT_2005, Taylor_abinitiotrans_2001}.

\subsection{Non-equilibrium Green's functions}
\label{NEG}

We have used the Keldysh formalism  \cite{Ryndyk_Green_2009} to compute the current through the central region.
In steady-state the two-time Green's function depends only on the time difference and can be treated efficiently in frequency space.
The steady-state current from the left transition layer to the central region can be calculated by integrating over the real part of the Keldysh Green's function
\begin{equation}
 I = \frac{e}{h} \int_{-\infty}^{\infty} d\omega \sum_{\begin{array}{c} i \in T_L \\ j\in C \end{array} } \Re \left( V_{ij}\mathcal{G}^{k}_{ij}(\omega) \right)\;.
\label{eq:current}
\end{equation}
Here, $i$ and $j$ denote the orbitals within the clusters $T_L$ and $C$, see \fref{fig:trans}, for calculating the current between the left transition layer and the central region.
In this paper we employ non-equilibrium cluster perturbation theory (n-CPT) \cite{Senechal_CPT_2008, nu.he.12}
to compute the Keldysh Green's function $\mathcal{G}$ 
of the interacting non-equilibrium system required in \eref{eq:current}.
Within n-CPT, \eref{eq:current} reduces to the well-known Landauer-B\"uttiker formula \cite{Meir_LandauerFormula_1992}.
\begin{equation}
 I = \frac{e}{h} \int_{-\infty}^{\infty} d\omega \left( f_R(\omega)-f_L(\omega) \right) T(\omega)
\end{equation}
$T(\omega)$ denotes the transmission function
\begin{equation}
 T(\omega) = \Tr \left( {\bf \Gamma }_{L} \bm{ \mathcal{G}}^{r}_{CC} {\bf \Gamma }_{R}  \bm{ \mathcal{G}}^{a}_{CC} \right) \;,
 \label{eq:T}
\end{equation}
where ${\bm \Gamma}_\nu$ with $\nu\in\{ L,R \} $ is the anti-hermitian part of the hybridization function
\begin{equation}
{\bm \Delta}^{r}_{\nu} = {\bf R}_\nu - \frac{i}{2}{\bm \Gamma}_\nu \;.
\end{equation}
We denote matrices with bold faced letters and for simplicity of the notation, we have omitted the $\omega$ arguments.
To this end, the entire system is decomposed into the device ($C$), e.g. the BDT molecule, and the two leads, which 
are further decomposed into
$L_{L}$ plus $T_{L}$ on the left side and $L_{R}$ plus $T_{R}$ on the right. 
We calculate the full Green's function for indices of the central region ${\bm{\mathcal{G}}}_{CC}$ within the CPT approximation, i.e.
we use the following Dyson equation
\begin{eqnarray}
 \big(\bm{{\mathcal{G}}}_{CC}^{r,a}\big)^{-1} &= \big({\bf g}_{CC}^{r,a}\big)^{-1} -\sum_{\nu\in\{ L,R \} }\underbrace{{\bf V}_{C T_{\nu}} {\bf G}_{T_{\nu}T_{\nu}}^{r,a}{\bf V}_{T_{\nu} C}}_{{\bf \Delta}_\nu^{r,a}} \nonumber \\ \label{eq:CPT}
 &= \big({\bf g}_{CC,0}^{r,a}\big)^{-1} -{\bf \Sigma }^{r,a} - \sum_{\nu\in\{ L,R \} }{\bf \Delta}_{\nu}^{r,a} \;.
\end{eqnarray}
In CPT the Green's function 
${\bf g}^{r,a}_{CC}$ of the isolated central cluster is required. 
From the  Dyson equation 
\begin{equation}\label{eq:}
\big({\bf g}^{r,a}_{CC}\big)^{-1} = \big({\bf g}^{r,a}_{CC,0}\big)^{-1} - {\bm \Sigma}^{r,a}
\end{equation}
it is clear that in the interacting case, the CPT approximation is equivalent to approximating 
the self-energy by that of the isolated cluster.
Therefore it is important to cut the system in a way that the coupling between the transition layers and the central region is 
weak compared to the hopping parameters within the central region.
The Green's function of the isolated central cluster can be calculated using the Lehmann representation
\begin{eqnarray}
 \left\{ {\bf g}^{r}_{CC}\right\}_{ij} = &\bra{\Omega} \opa_i \frac{1}{\omega + i0^+ -\opH_C + E_0} \opad{j} \ket{\Omega} \nonumber\\
 &+ \bra{\Omega} \opad{j} \frac{1}{\omega + i0^+ +\opH_C - E_0} \opa_i \ket{\Omega}\;.
 \label{eq:Lehmann}
\end{eqnarray}
If the system is not too far away from the equilibrium situation, it will still be a good approximation to consider only excitations from the ground state to higher states
and hence the corresponding self-energy.

The Green's function of the 
isolated left lead with indices restricted to the transition region ${\bf G}_{T_{L}T_{L}}$ is needed in \eref{eq:CPT}.
We will also need ${\bf G}_{T_{R}T_{R}}$.
To calculate ${\bf G}_{T_{\nu}T_{\nu}}$ for $\nu\in\{L,R\}$ we once more introduce a
cluster decomposition of  the leads into part $L_{\nu}$ and $T_{\nu}$, introduced before.
The exact Dyson equation for this decomposition yields
\begin{equation}
 \left( {\bf G}_{T_{\nu}T_{\nu}}\right) ^{-1} = \left( {\bf g}_{T_{\nu}T_{\nu}}\right) ^{-1} - {\bf V}_{T_{\nu} L_{\nu}}\; {\bf g}_{L_{\nu} L_{\nu}} \; {\bf V}_{L_{\nu} T_{\nu}}.
\end{equation}
where  ${\bf g}_{T_{L}T_{L}}$ stands for the Green's function of the isolated  transition region, which can be computed easily, as the matrix dimension is small.
${\bf g}_{L_{\nu}L_{\nu}}$ on the other hand represents the Green's function of the isolated semi-infinite part $L_{\nu}$
of lead $\nu$ that is composed of identical unit cells.
The hopping elements ${\bf V}_{T_{\nu},L_{\nu}}$ couple only to the unit cell near to the transition region in each lead and therefore only the surface part of ${\bf g}_{L_{\nu}L_{\nu}}$ is required, which can be determined iteratively by using the method of Sancho et al. 
\cite{Sancho_tranfermatrix_1984}. \rumcomm{An alternative technique for calculating the hybridization function is to use complex absorbing potentials \cite{Driscoll_absorbingpotential_2008}.}

A bias voltage $V_b$ drives the system out of equilibrium and enters the calculation as a shift of the on-site energies and the chemical potential by $-V_b/2$ for the left lead and by $+V_{b}/2$ for the right lead.
In the Landauer-B\"uttiker formula $V_b$ enters also in the Fermi functions.
A gate voltage $V_g$ can be applied by shifting the on-site energies of $\tilde{t}_{\bar{i}{\bar{j}}}$.

The spectral density and the conductivity are important quantities for our analyses and they are defined as
\begin{eqnarray}
 A_{\nu\nu}(\omega) &= -\frac{1}{\pi} \Im \Tr \left({\bf g}^{r}_{\nu\nu}(\omega) \right) \\
 G &= \frac{d I}{d V_b} \;.
\end{eqnarray}

\section{Applications}
\label{Applications}

\subsection{The Au-BDT-Au system}

In a first step we have applied the DFT+NEGF formalism to a system consisting of a BDT molecule sandwiched between two gold chains, schematically drawn in \fref{fig:trans}.
As shown below  in the case of monoatomic gold chains it is possible to include all lead atoms in the Wannier transformation in a controlled way.
In mechanically controllable break junctions (MCBJ) or with an scanning tunnelling microscope (STM) the formation of monoatomic gold chains was demonstrated under elongation of the junction by several authors \cite{French_AuMACsim_2013, Quian_quasi1delectrodes_2007, Coura_AuMAC_2004, Yanson_AuMAC_1998, Ohnishi_AuMAC_1998}.
In a first step we have performed DFT 
calculations~\footnote{We used a cut-off energy for the wavefunction of $E_{\rm cowf} =64~$Ry and the charge density of $E_{con}=512~$Ry,
the Methfessel-Paxton smearing technique with the smearing parameter $E_{\rm tsm}=0.01~$Ry,
the vacuum distance $d_{\rm vac}=12~$\AA~
and $32$ k-points.} for the periodic chain of monoatomic gold atoms.
All the parameters are converged with respect to the total energy within $5~$meV.
In order to find the equilibrium geometry, we have calculated the total energy of the chain for different distances $d_{Au}$ between the gold atoms.
We find an optimal distance of $d_{Au} = 2.515~$\AA~ which is close to the LDA result $2.51~$\AA~ in \cite{Sclauzero_Auchain_2012}.
\begin{figure}[t]
\includegraphics[width=1.0\columnwidth,angle=0]{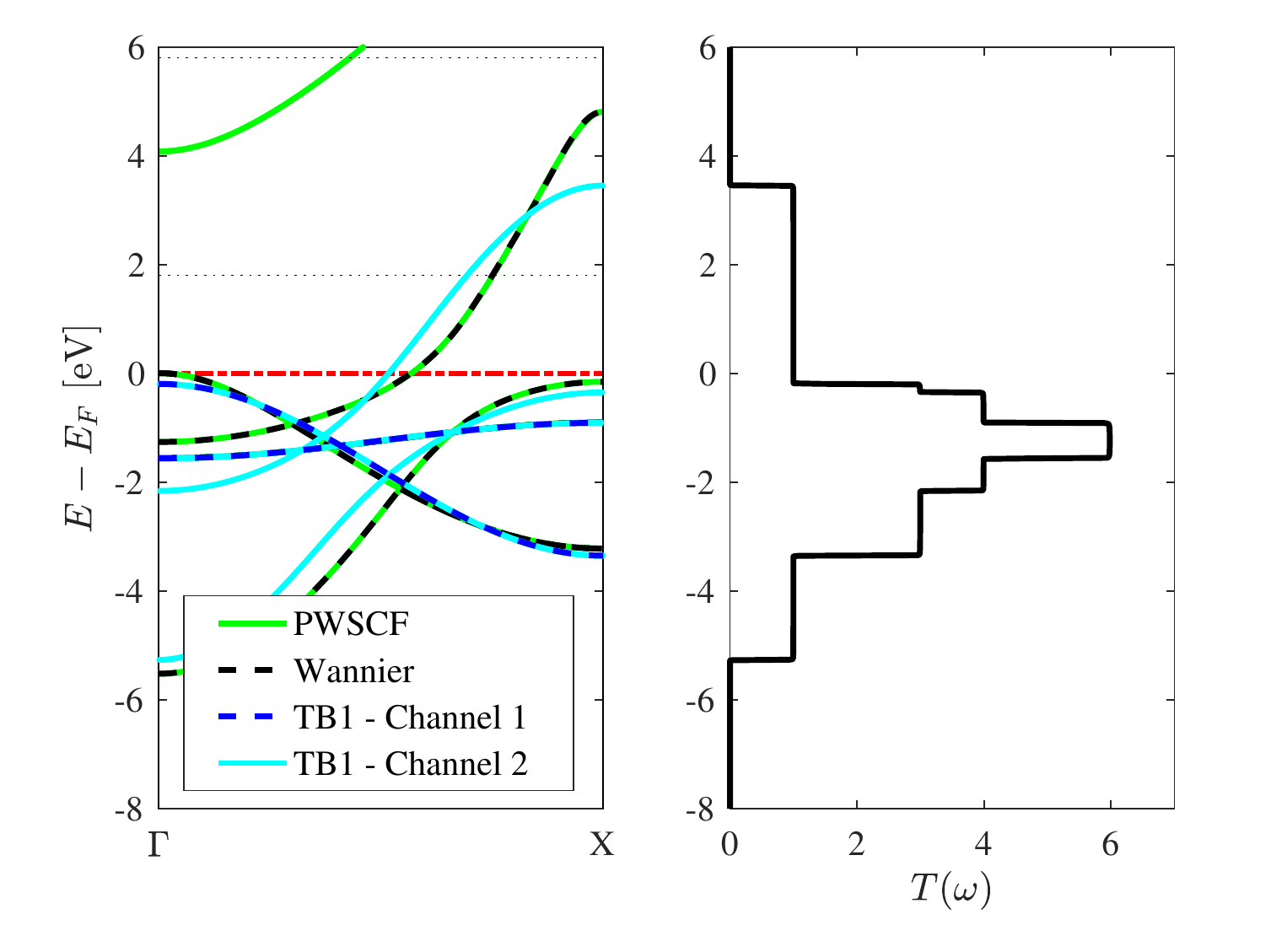}
\caption{Construction of the single-particle Hamiltonian for the monoatomic gold chains. Left panel: plane wave band structure (PWSCF), Wannier bands (Wannier), nearest neighbour tight binding bands (TB1). Right panel: transmission function, calculated from TB1.
}
\label{fig:Auchain}
\end{figure}

\begin{figure}[t]
\includegraphics[width=1.0\columnwidth,angle=0]{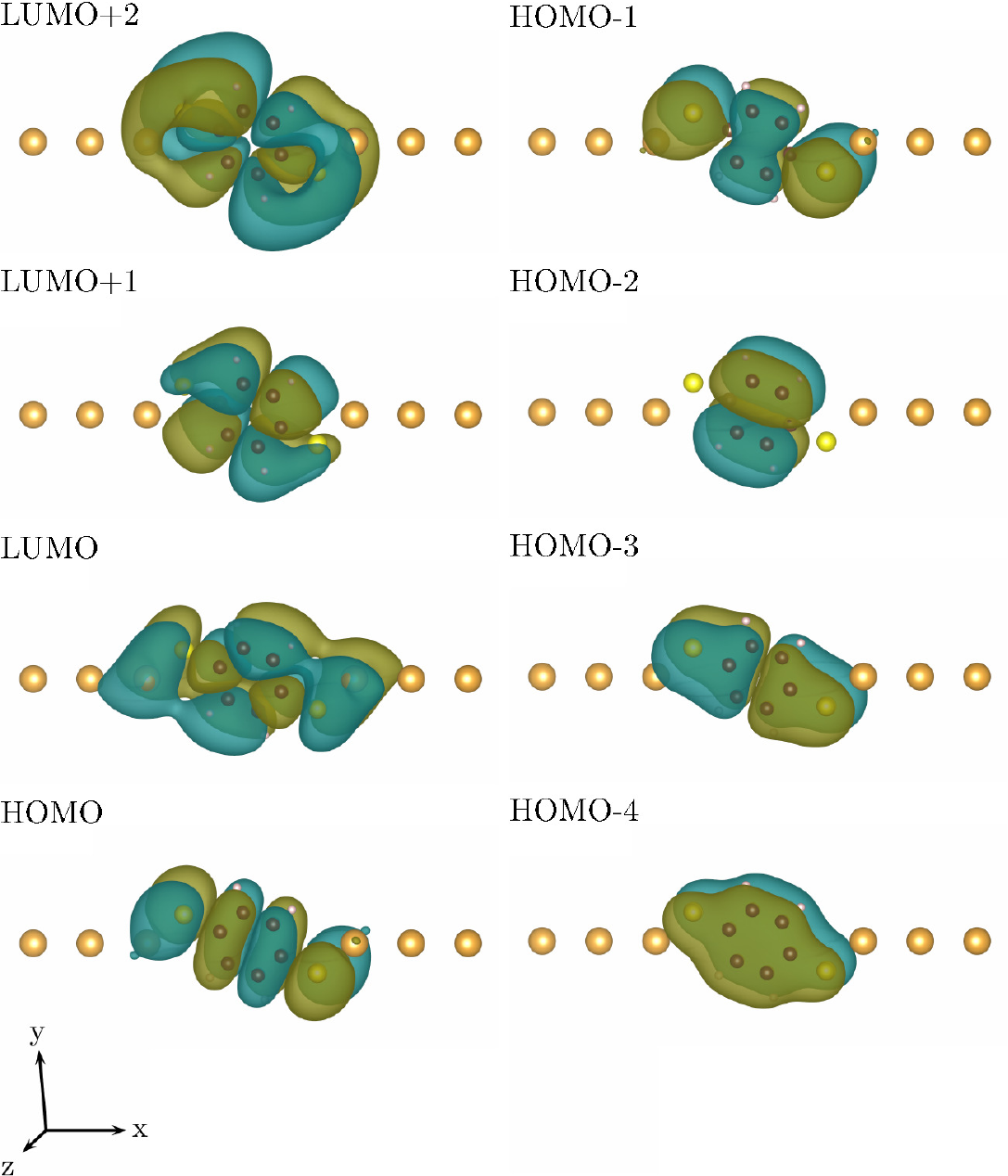}
\caption{Eigenorbitals of channel 1 in the central region of the Au-BDT-Au transport system represented with VESTA \cite{Momma_VESTA_2011}.}
\label{fig:AuchIEig}
\end{figure}

The band structure of the Au chain in an interval around $E_F$ is shown in the left panel of \fref{fig:Auchain}.
The $\Gamma$-$X$ axis is always the axis along the gold chain in reciprocal space.
The 6p bands are at higher energies.
The 6s and 5d bands coincide exactly with the Wannier bands, which demonstrates the properness of the Wannier transformation.
The dotted lines represent the inner and the outer energy window needed to disentangle the 6s and 5d bands from the 6p band in the Wannier transformation.
In \fref{fig:Auchain} tight binding (TB1) means that only  hopping processes  to nearest neighbour gold atoms are taken into account.
The 6s band and the 5d bands separate into the two bands corresponding to the $d_{xz}$ and the $d_{yz}$ orbital (Channel 1)
and four bands corresponding to orbitals with inversion symmetry with respect to the $xy$-plane (Channel 2).
The definition of the coordinate system is according to \fref{fig:AuchIEig}.
Each symmetry channel will only couple to orbitals of the same symmetry and therefore transport 
takes place in two separated channels.
The TB1 approximation turns out to be a good approximation for the channel 1 bands, which are mainly interesting for transport.
The right panel of \fref{fig:Auchain} is the calculated equilibrium ($V_b=0$) transmission function for TB1 and is similar to the results obtained in \cite{Sclauzero_Auchain_2012}.
In the coherent regime the transmission function is proportional to the number of bands at a certain energy. 

\begin{figure}[t]
\includegraphics[width=1.0\columnwidth,angle=0]{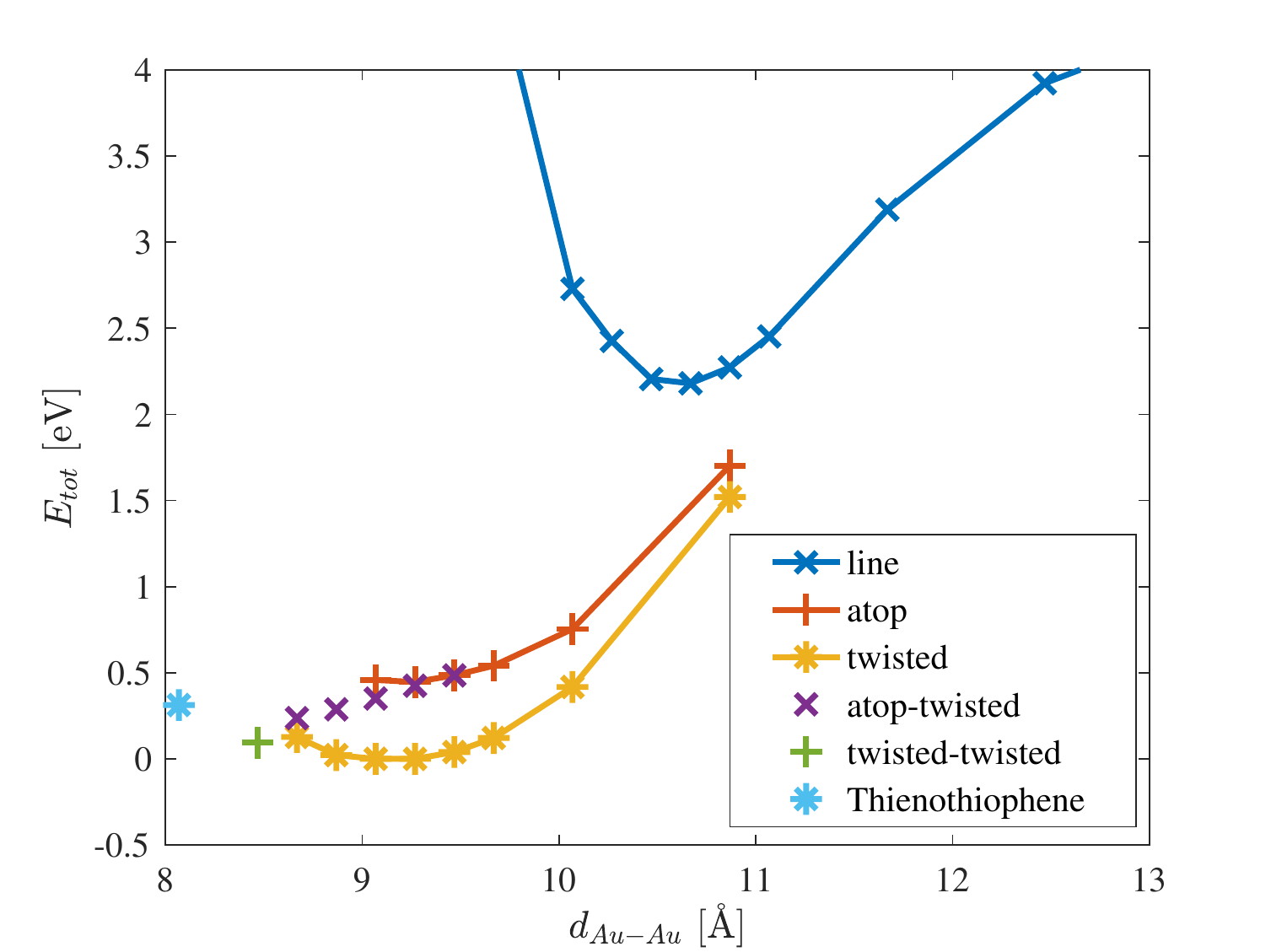}
\caption{Total energy of the the Au-BDT-Au system as function of the distance between neighbouring gold atoms and for different geometries. The total energy minimum is set to be at $0~$eV.}
\label{fig:PhaseDiagram}
\end{figure}

After having calculated the Hamiltonian of the monoatomic Au chain, we considered the full system, including BDT \footnote{We used a cut-off energy for the wavefunction of $E_{\rm cowf} =100~$Ry and the charge density of $E_{con}=400~$Ry,
the Methfessel-Paxton smearing technique with the smearing parameter $E_{\rm sm}=0.001~$Ry,
the vacuum distance $d_{\rm vac}=18~$\AA~
and $1$ k-point.}.
The BDT molecule can assume different geometry configurations between the two semi-infinite gold chains.
In the \textit{line} configuration the BDT molecule is forced to be on the gold chain axis.
In the \textit{atop} configuration the BDT molecule slightly hops out of the gold chain axis, while
it twists out of the axis in the \textit{twisted} configuration as shown in \fref{fig:trans}.
There are also mixed configurations between atop and twisted, called \textit{atop-twisted} and \textit{twisted-twisted}.
In order to determine the  energetically favourable configuration, we compute
the total energy as function of the distance between the gold atoms near to the BDT molecule $d_{Au-Au}$, {namely the gap between the two (left and right) semi-infinite gold chains},  
for different configurations, see \fref{fig:trans} and \ref{fig:PhaseDiagram}.
The distance between the gold atoms was fixed $d_{Au}$ during the geometry optimizations.
At $d_{Au-Au} \approx 8.1$ the BDT molecule rearranges into thienothiophene.
It occurs  that the twisted configuration of the BDT atom is energetically most favoured and, consequently,  we have chosen this configuration for the ensuing transport calculations.
The optimized distance is $d_{Au-Au} = 9.16~$\AA.
We have chosen 6 gold atoms per unit cell, which ensures that the eigenvalues of the atoms at the outer edges of the unit cell (see \fref{fig:trans})
are converged. We have chosen the size such that the changes in eigenenergies are less than $0.05~$eV.
\begin{figure}[t]
\includegraphics[width=1.0\columnwidth,angle=0]{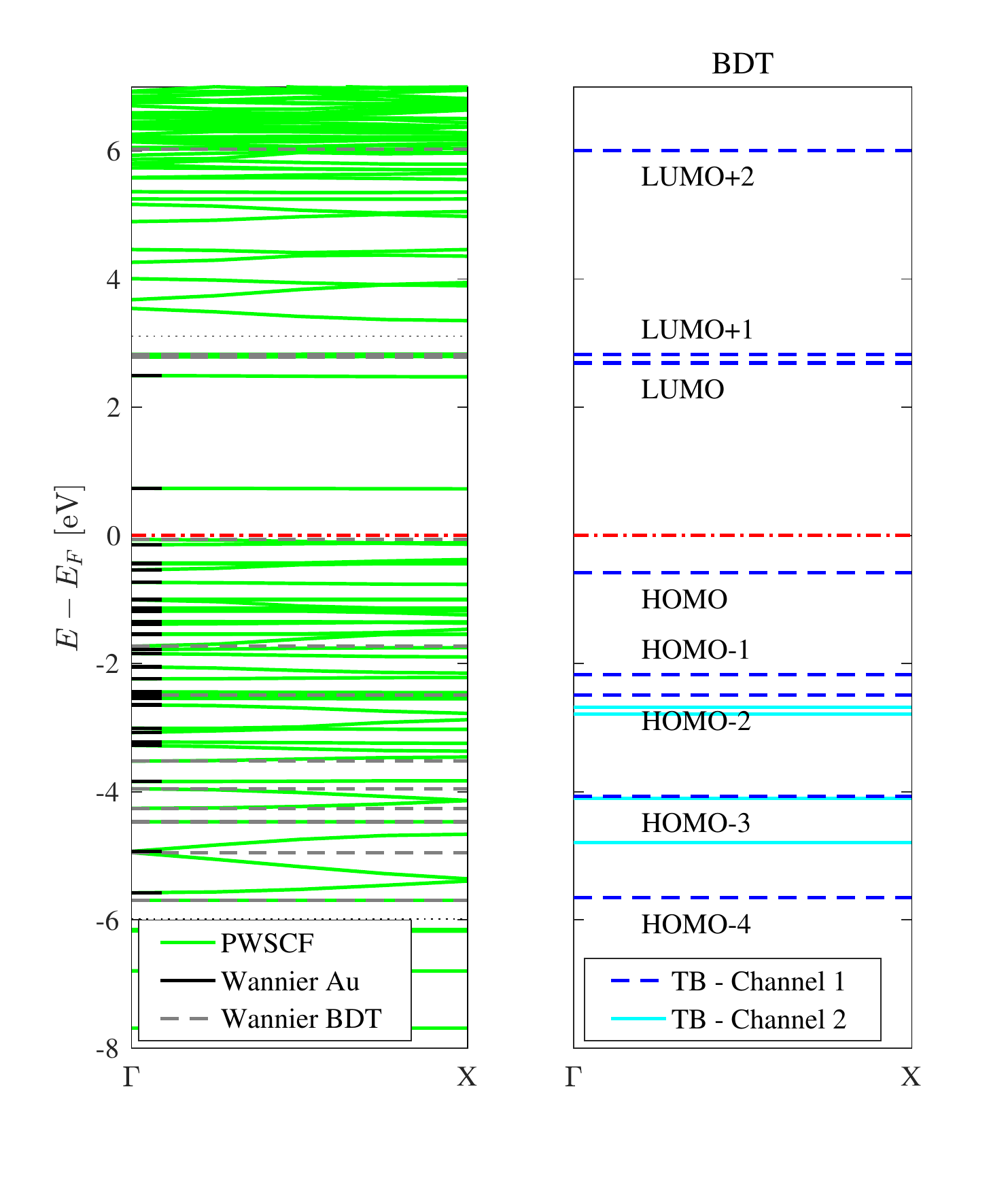}
\caption{The plane wave band structure (PWSCF) and the corresponding Wannier transformation (Wannier) (left panel). The Wannier bands are projected onto onto orthogonalized atomic wavefunctions to show whether they correspond to the gold atoms (Au) or the BDT molecule (BDT).
The levels of the decoupled BDT molecule split into channel 1 and channel 2 (right panel).}
\label{fig:WanAu}
\end{figure}
The left panel in \fref{fig:WanAu} depicts a comparison of the band structure obtained with the full plane wave basis and with the Wannier basis.
The Wannier transformation is performed only at the $\Gamma$-point.
In the Wannier transformation we picked out 47 bands, 35 corresponding to the 6 gold atoms and there are 12 BDT bands within this energy range.
At the $\Gamma$-point the PWSCF bands coincide exactly with the Wannier levels, which demonstrates again the properness of the Wannier transformation.
The obtained Wannier basis functions of the gold atoms coincide with the Wannier basis functions of the pure gold chain, which shows that the disentanglement procedure of the 6s and 5d bands from the 6p bands is done correct.
Projections of the wavefunctions onto orthogonalized atomic wavefunctions
show whether the Wannier levels belong to the gold leads and the transition layers or to the BDT molecule.
The right panel presents the eigenlevels of the decoupled BDT molecule, the eigenenergies of $\opH_C$.
As mentioned above and outlined in \cite{Ventra_BDT_2000, Kondo_BDTchannel_2006} 
transport takes place in separated channels.
Channel 1 (blue dashed) consists of $p_z$-like BDT orbitals 
and channel 2 (cyan) of $p_{xy}$-like BDT orbitals.
Each channel couples only to the corresponding channel in the gold leads.

The eigenorbitals of channel 1 are shown in \fref{fig:AuchIEig}.
The shape of the orbitals can be explained by the symmetry of the benzene molecule, which has the point group $D_{6h}$.
LUMO+2 is the $B_{2g}$-orbital, LUMO+1 and LUMO are the $E_{2u}$-orbitals, HOMO-4 is the $A_{2u}$-orbital and the others are a mixture of the benzene $E_{1g}$- and the sulfur $p_z$-orbitals.
%
Due to the fact that only the HOMO level is close to the Fermi energy it is the level mainly contributing to transport which is in general the case in systems with thiol-anchoring groups \cite{Zotti_SingleLevelModel_2010}.
The HOMO orbital is an unsaturated sulfur $p$-orbital and can also be compared to \cite{Rangel_PDOStrans_2015, Souza_thiolthiolate_2014}.
Passivating the unsaturated sulfur $p$-orbitals would shift the HOMO level away from the Fermi energy and produce lower conductances \cite{Thygesen_BDT_2011}.
\rumcomm{Charge transport through channel 1 is blocked due to the fact that there are no empty states with $p_z$-like symmetry in the monoatomic gold chain (only a $sd$-hybridized orbital belonging to channel 2 is above the Fermi level in \fref{fig:Auchain}).
The levels corresponding to channel 2 are more than $2~$eV below the Fermi energy.
Therefore,  transport in this channel is due to the broadening of the leads and
the current-voltage characteristic nearly linear.
The conductivity for the BDT molecule connected with monoatomic Au chains is almost constant and in the order of $0.01 ~G_0$ for biasvoltages bellow $4~$V which
is in the order of the experimental values of $0.01~G_0$ \cite{Song_exp_2009, Xiao_exp_2004, Tsuitsui_exp_2009, Kiguchi_exp_2010}.
But using gold tips or bulk-like gold as leads \cite{Kondo_BDTchannel_2006, Ventra_BDT_2000} produces empty states with $p_z$-like symmetry and activates channel 1 which raises the conductance.
Therefore the theoretical LDA result for the Au-BDT-Au system with bulk-like leads is $0.28~G_0$ \cite{Thygesen_BDT_2011}.

Since channel 1 contains the HOMO-orbital and the typical benzene $p_z$ orbitals it would be the most important channel if supported by the leads.
The coupling between leads and the channels depends on the geometry of the junction and therefore as mentioned above gold tips or bulk-like gold as leads \cite{Kondo_BDTchannel_2006, Ventra_BDT_2000} activate channel 1.
Also the use of platinum instead of gold opens channel 1 because
platinum has one electron less than gold and therefore also orbitals with $p_z$-like symmetry contribute to transport.}

\subsection{The Pt-BDT-Pt system}

\begin{figure}[t]
\includegraphics[width=1.0\columnwidth,angle=0]{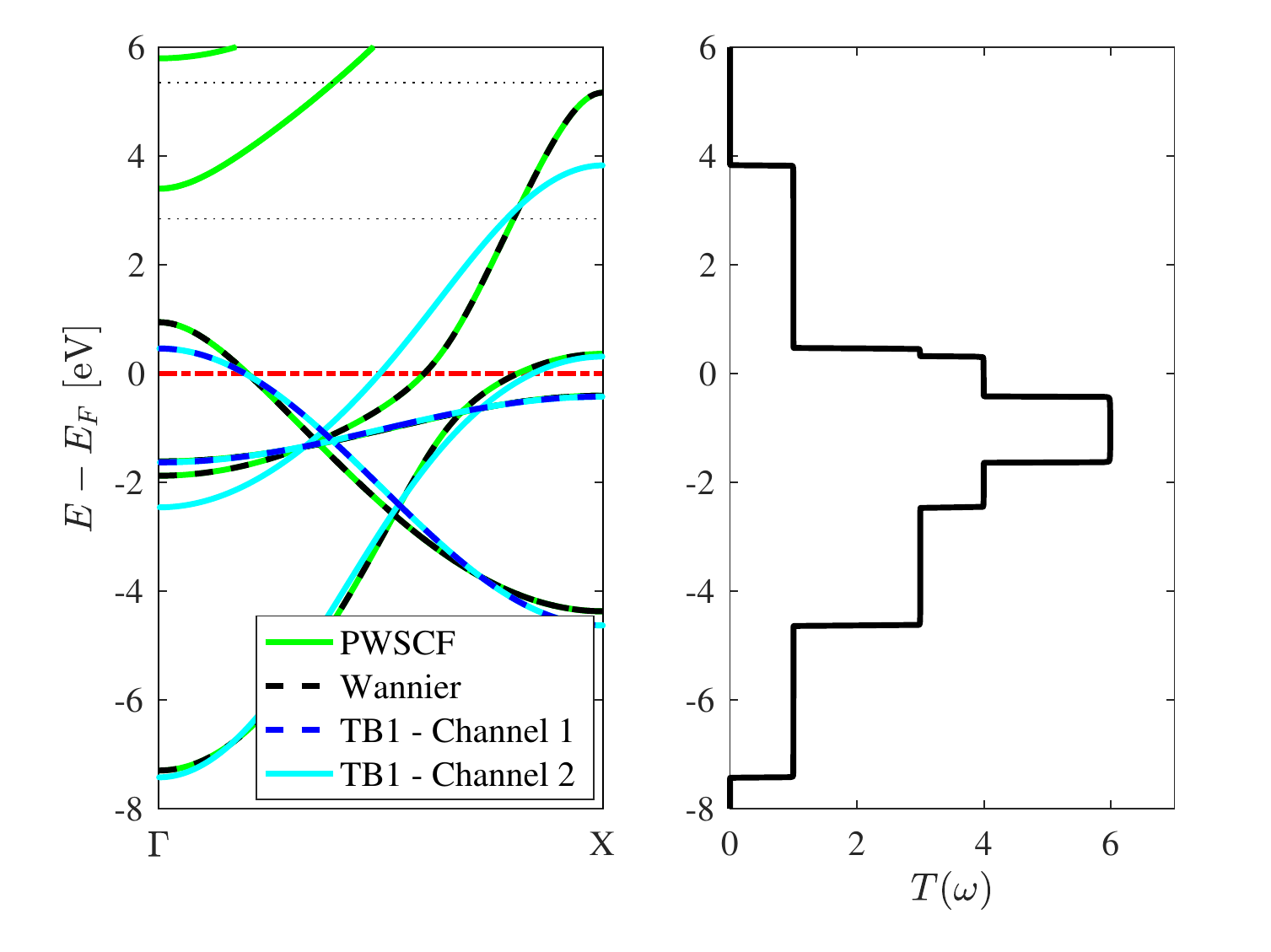}
\caption{Construction of the single-particle Hamiltonian for the monoatomic platinum chains. Left panel: plane wave band structure (PWSCF), Wannier bands (Wannier), nearest neighbour tight binding bands (TB1). Right panel: transmission function, calculated from TB1.
}
\label{fig:Ptchain}
\end{figure}

In order to activate channel 1 we have replaced the gold atoms by platinum ones in our calculations \footnote{We used a cut-off energy for the wavefunction of $E_{cowf} =60~$Ry and the charge density of $E_{con}=200~$Ry,
the Methfessel-Paxton smearing technique with the smearing parameter $E_{sm}=0.005~$Ry,
the vacuum distance $d_{vac}=12~$\AA~
and $60$ k-points.}.
The optimized distance between the platinum atoms $d_{Pt} = 2.328~$\AA~ is comparable to the value of $2.34~$\AA~ in \cite{Sclauzero_Ptchain_2008}.
The plane-wave pseudopotential band structure, Wannier bands and TB1 bands along with the transmission function are shown in \fref{fig:Ptchain}.
In the Pt chain not just channel 2 but also channel 1 bands contribute to the equilibrium transmission at the Fermi energy.

\begin{figure}[t]
\includegraphics[width=1.0\columnwidth,angle=0]{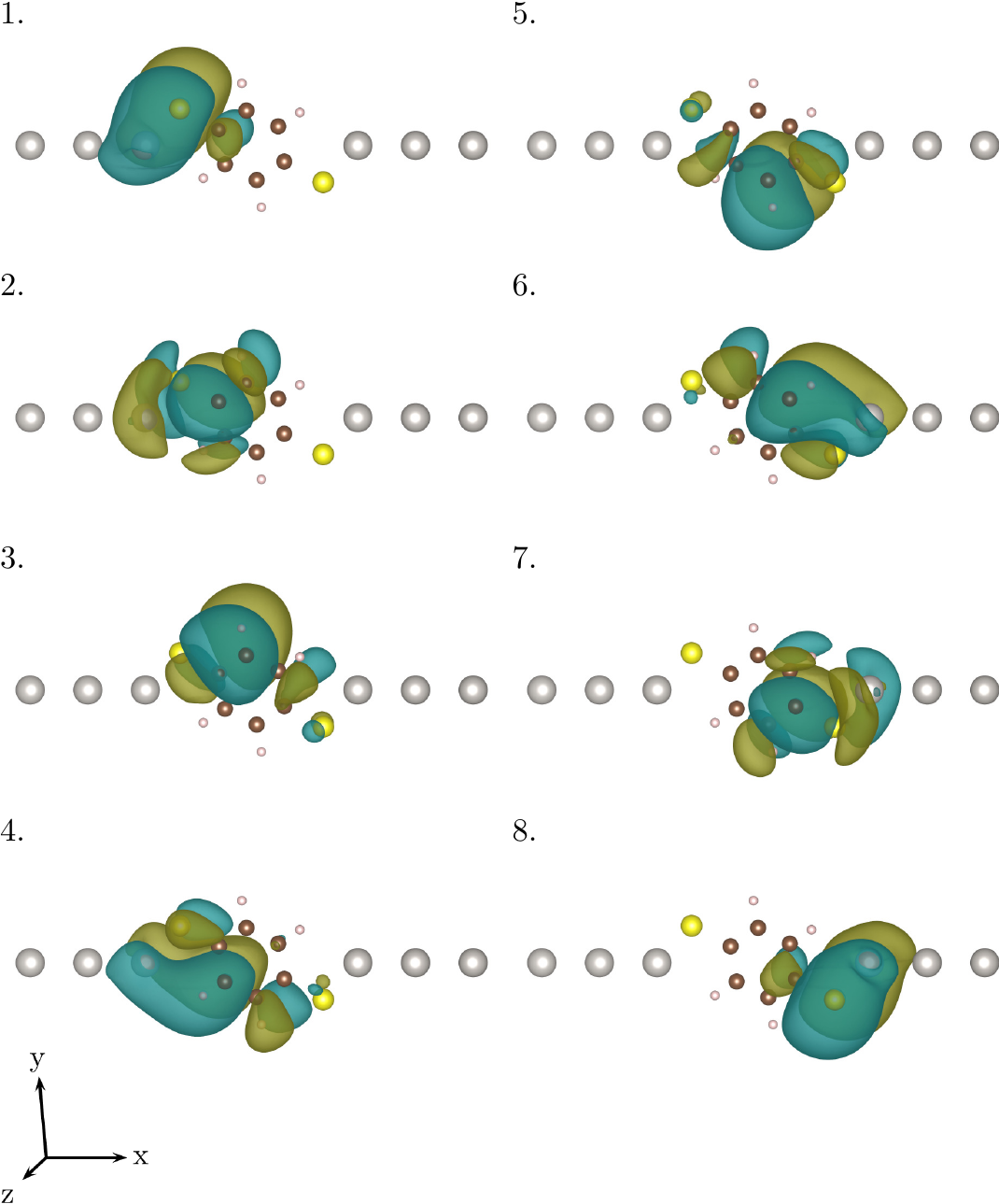}
\caption{Localized molecular orbitals of channel 1 in the central region of the Pt-BDT-Pt transport system represented with VESTA \cite{Momma_VESTA_2011}.}
\label{fig:PtchIWan}
\end{figure}

We now consider the Pt-BDT-Pt system \footnote{We used a cut-off energy for the wavefunction of $E_{cowf} =80~$Ry and the charge density of $E_{con}=320~$Ry,
the Methfessel-Paxton smearing technique with the smearing parameter $E_{sm}=0.005~$Ry,
the vacuum distance $d_{vac}=15~$\AA~
and $1$ k-point.}.
\begin{figure}[t]
\includegraphics[width=1.0\columnwidth,angle=0]{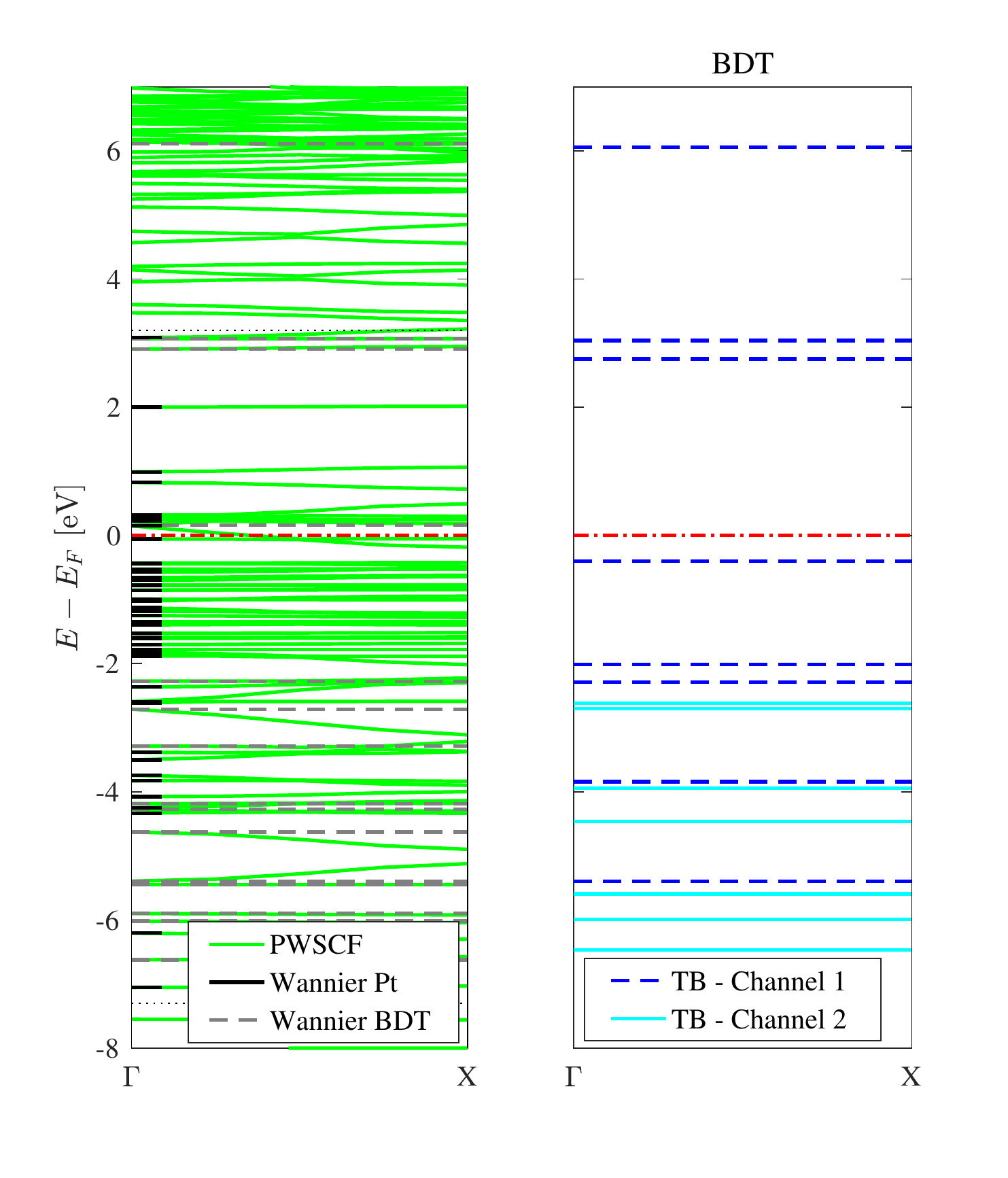}
\caption{
The plane wave band structure (PWSCF) and the corresponding Wannier transformation (Wannier) (left panel). The Wannier bands are projected onto onto orthogonalized atomic wavefunctions to show whether they correspond to the platinum atoms (Pt) or the BDT molecule (BDT).
The levels of the decoupled BDT molecule split into channel 1 and channel 2 (right panel).}\label{fig:WanPt}
\end{figure}
The optimized distance between the Pt atoms near to the BDT molecule is $d_{Pt-Pt} = 8.58~$\AA.
It is necessary to take at least 8 platinum atoms per unit cell to ensure that the influence of the BDT molecule on the lead platinum atoms is negligible.
The comparison of the PW band structure with the Wannier bands is given in the left panel of \fref{fig:WanPt}.
In the Wannier transformation we retained 62 bands, 47 corresponding to the 8 platinum atoms and there are 15 BDT bands within this energy range.
The right panel presents the decoupled levels of the BDT molecule and is comparable to \fref{fig:WanAu} disregarding 3 more levels in channel 2 needed to get all the platinum levels in the calculation.
In the following we will only discuss channel 1 because due to the level alignment the current in channel 2 is an order of magnitude smaller.
The orbitals of channel 1 look similar to the ones in the Au-BDT-Au calculation in \fref{fig:AuchIEig}.
The orbitals in \fref{fig:PtchIWan} are the localized Wannier orbitals of the BDT molecule in the Pt-BDT-Pt calculation. 
These are the 8 orbitals entering the TB model for the central region.
They all have $p_z$ symmetry.
The matrix form of the TB Hamiltonian of the central region, in the basis of the localized orbitals in \fref{fig:PtchIWan}, and the coupling matrices to the leads that have been used in the calculations are given in \sref{parameters}.
The Fermi energy in the transport Hamiltonian is adjusted such that the local DOS of channel 1 of the atom further away from the BDT molecule has the same filling as in the pure atom chain.

\begin{figure}[t]
\includegraphics[width=1.0\columnwidth,angle=0]{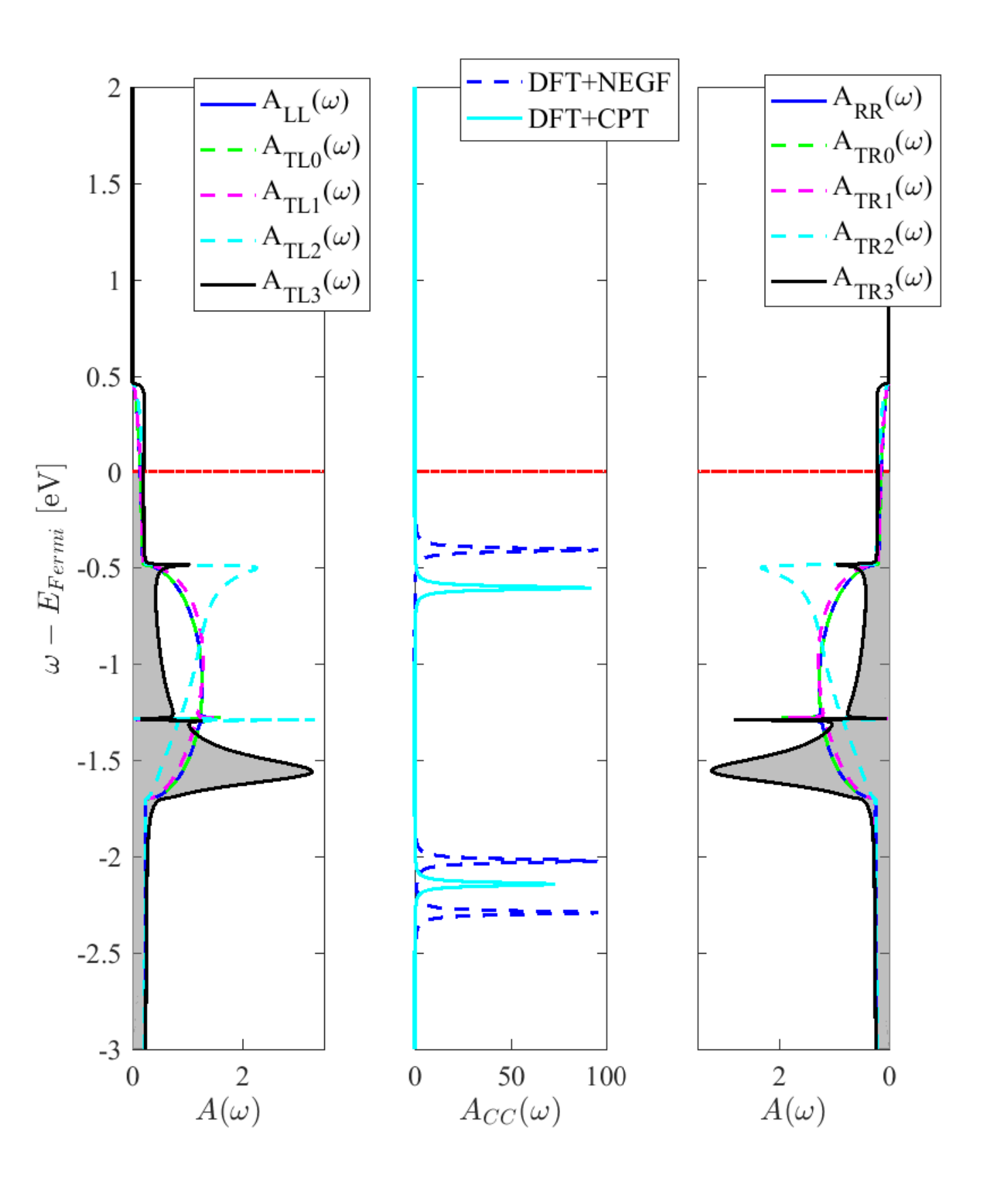}
\caption{The coloured lines represent the spectral function at different points in the transition region (first and third panel). 
 The numbers in the indices stand for the distance of these points from the outermost edges of the transition region. The grey shaded area indicates the filling.
The spectral function of the center is drawn in the middle panel.}
\label{fig:transport}
\end{figure}

The plot in the middle of \fref{fig:transport} shows the spectral density $A_{CC}(\omega)$ of the isolated central cluster, namely the isolated BDT molecule.
The curves labelled by DFT+CPT represent the result for the correlated system, which will be explained in \sref{manybody}.
The peaks in the DFT+NEGF result are at $-0.4~$eV, $-2.0~$eV and $-2.3~$eV.
As discussed above the HOMO level closer to the Fermi level ($-0.4~$eV) is mainly responsible for the transport.
The spectral density at  the surface atom of the left and the right lead is shown in the left and the right panel of \fref{fig:transport}.
The spectral density of a homogeneous multiorbital chain is a superposition of semi-circles, see $A_{LL}(\omega)$ and $A_{RR}(\omega)$.
Coupling  the homogeneous chains to the the transition layers  changes the semi-circular structure of the spectral density.
The spectral densities $A_{TLi}(\omega)$ and $A_{TRi}(\omega)$ are located $i$ atoms away from the last point of the transition region, i.e. points with larger $i$ 
are closer to the central regions. We nicely see, how the spectral function 
gradually changes from the multi-circular structure of the homogeneous chain to the 
structure of the spectral densities $A_{TL3}(\omega)$ and $A_{TR3}(\omega)$, which consists of 
peaks at $-1.6$ and $-0.5$ and a spike at about $-1.3$.
The grey shaded area indicates the filling. The remarkable point is that the spectral function of the transition region, that enters the transmission function and therefore the current,
has a complex structure that will even change when a voltage is applied.
\rumcomm{As demonstrated by Cuniberti et al. \cite{Cuniberti_leads_2001, Cuniberti_Contacts_2002} low dimensional leads can affect the conductance due to the finite band width and the structure in the DOS.
 If the DOS is nearly constant one can use the wide-band limit (WBL) approximation which works well for systems with bulk metal electrodes \cite{Verzijl_WBL_2013}.
 However the DOS of a one-dimensional chain has Van Hove singularities near the band edge and the WBL approximation does not work.}
The key message therefore is that the details of the leads in such a molecular device are of crucial importance for the transport properties.
This adds to the observation that the symmetry of the molecular orbitals and the leads can lead to selection rules as far as transport channels are concerned.

\subsection{Stability diagrams of the Pt-BDT-Pt system}

\begin{figure}[t]
\includegraphics[width=1.0\columnwidth,angle=0]{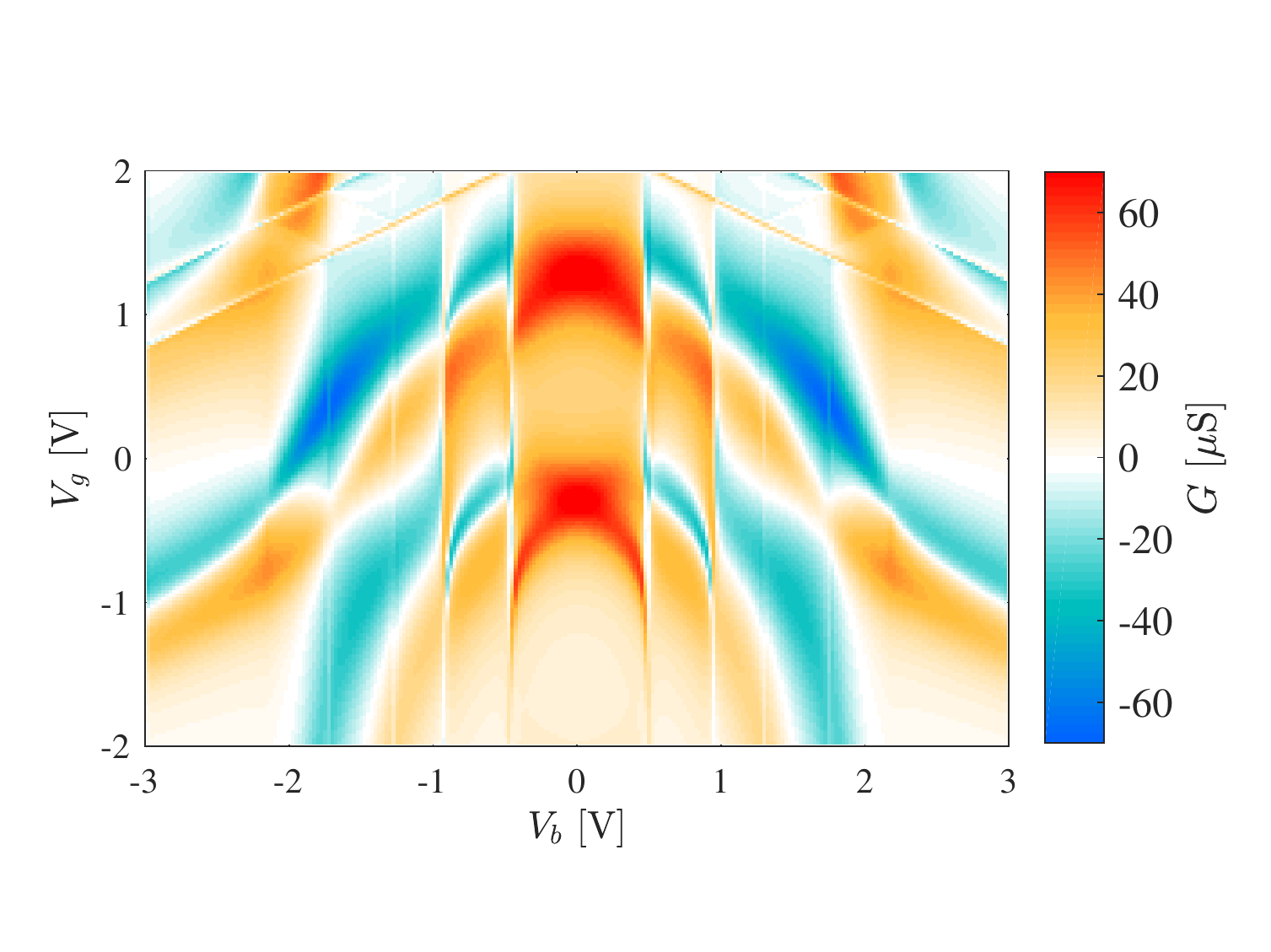}
\caption{Charge stability diagram obtained from the LDA data.}
\label{fig:currentLDA}
\end{figure}

Next we present various aspects of the charge stability of the Pt-BDT-Pt system. 
The stability diagram computed with the DFT+NEGF formalism is presented in \fref{fig:currentLDA}.
We have restricted the bias voltage to  $-3~$V$ < V_b < 3~$V, because otherwise we would have to include platinum $p$-orbitals, and likewise we have restricted the gate voltage to the interval $-2~$V$ < V_g < 2~$V,  otherwise channel 2 becomes important.
All diagrams are calculated at an inverse lead temperature of $\beta = 300~1/$eV, which enters in the Fermi functions of the leads.
\rumcomm{The conductance at $V_b=V_g=0$ is $0.52~G_0$ and is considerable higher than the experimental ($0.01~G_0$ \cite{Song_exp_2009, Xiao_exp_2004, Tsuitsui_exp_2009, Kiguchi_exp_2010}) and theoretical LDA ($0.28~G_0$ \cite{Thygesen_BDT_2011}) values of Au-BDT-Au systems.
Beside the influence of the different contact material the formation of monoatomic chains results in large increases in the conductance \cite{Souza_thiolthiolate_2014, French_AuMACsim_2013, Sergueev_BDTElongation_2010}.
The distance $d_{Pt-Pt}$ is also small compared to experimental values, where the monoatomic chains arise under applying stress to the leads.
This results in a higher coupling strength between the leads and the central region.}

Most strikingly, the stability diagram in \fref{fig:currentLDA} does not show the usual structure of
crossing straight lines, resulting in rhombic patterns.
The only such structure is the pair of straight lines starting at 
the left edge of the diagram with $V_{g}= 0.8$ and
$V_{g}= 1.3$, respectively. They correspond  to the level at $-2.3~$eV. At $V_{b}=3$
the Fermi energy of the right leads is shifted to $\mu=-1.5$. Above the Fermi level 
the remaining DOS of unoccupied states has a width of $\Delta E_{empty} = 0.5$ eV, see \fref{fig:transport}. Hence, the transport window of the right lead is $(-1.5,-1.0)$~eV. With $V_{g}=0.8$ the level, which was
originally at $-2.3$~eV is shifted to the lower edge of the transport window and 
with $V_{g}=1.3$ it is shifted to the upper edge.
This constant distance between the lines is an effect of the band edge
according to which a positive conductance is always followed by a negative one.
In our case the distance between the lines along the $V_b$-axis is $\Delta V_b = 2\Delta E_{empty}=1~$eV.
The factor two results from the fact that half of the bias voltage is applied to each lead.
The slope of the lines is constant and equal to $\pm 1/2$. If the leads are shifted with $\pm V_b/2$, one needs  $2 V_b$ to compensate a shift in $V_g$.
As a consequence of weak coupling of this level to the leads the effects can be seen so clearly.

\begin{figure}[t]
\includegraphics[width=1.0\columnwidth,angle=0]{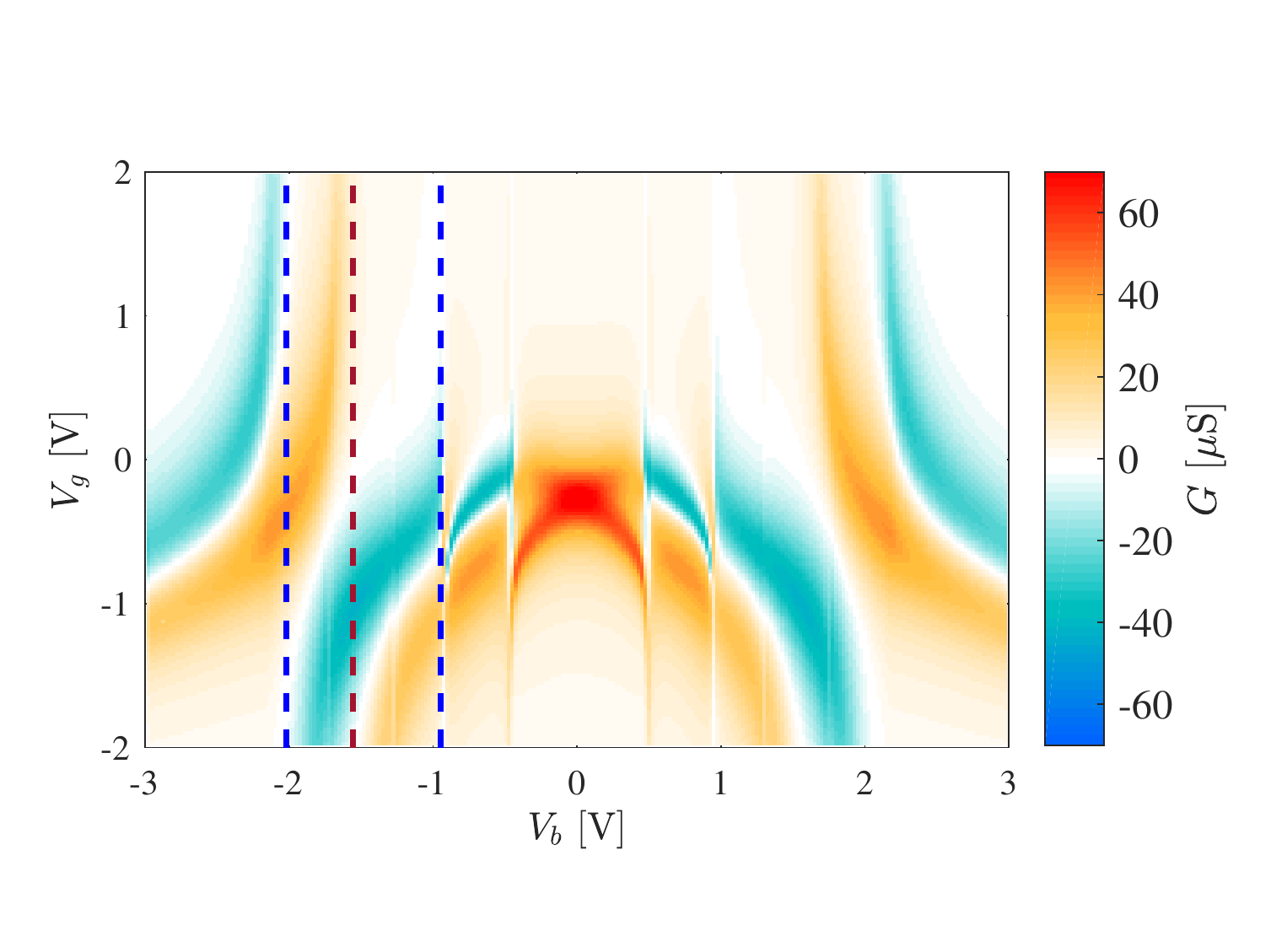}
\caption{Stability diagram of the LDA calculation corresponding to the level at $-0.4~$eV.}
\label{fig:SLmodel}
\end{figure}

\begin{figure}[t]
\includegraphics[width=1.0\columnwidth,angle=0]{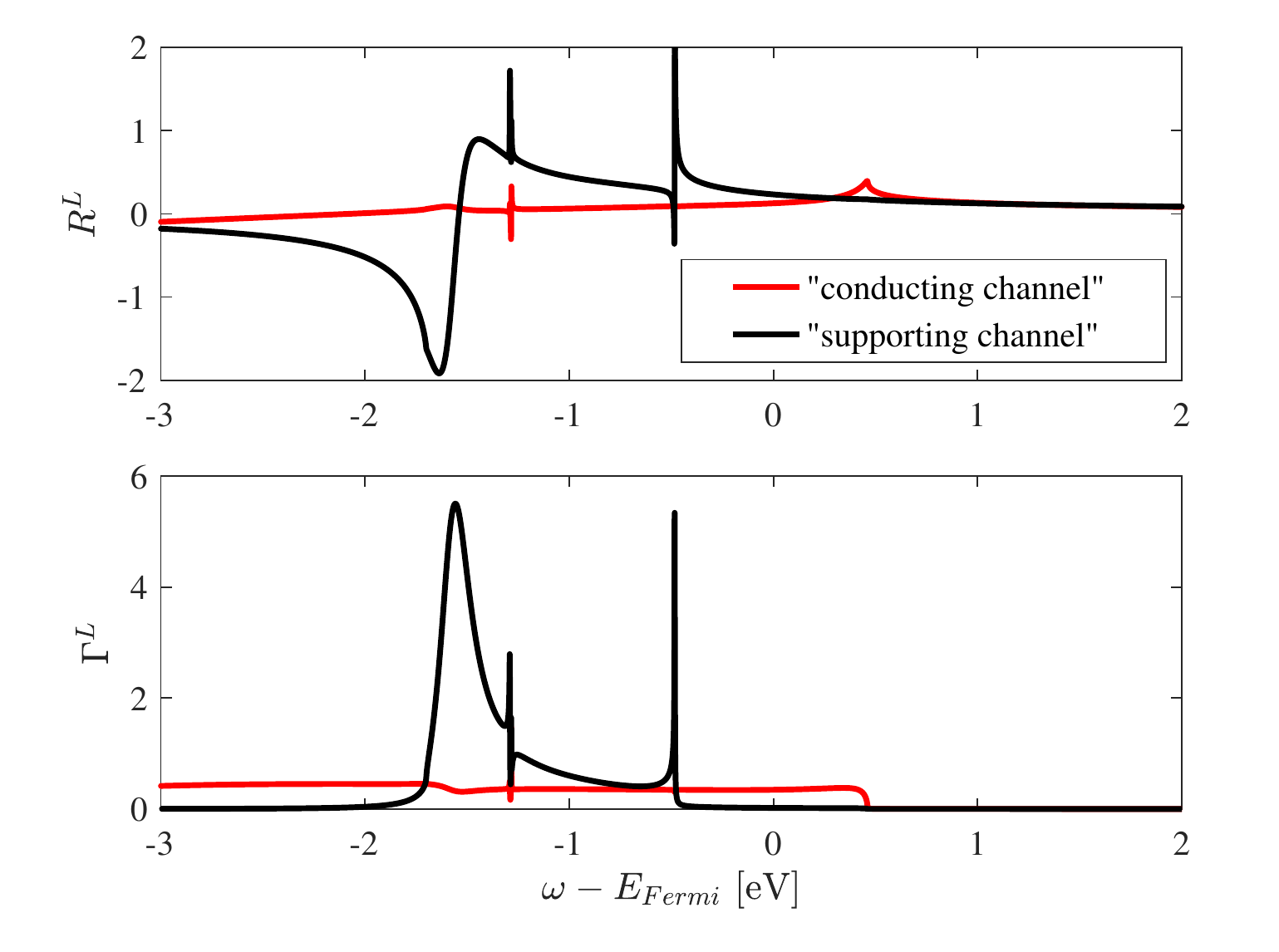}
\caption{Real $R^L$ and imaginary part $\Gamma^L$ of the retarded left lead hybridization function $\Delta^{L,r}$.}
\label{fig:sdt}
\end{figure}

The levels at $-2.0~$eV and $-0.4~$eV produce a similar structure but at different energies.
The structure can be explained by a \quot{band edge effect} and a \quot{supporting effect}.
To explain the \quot{supporting effect} we have projected out the level at $-0.4~$eV.
The resulting charge stability diagram is depicted in \fref{fig:SLmodel}.
Comparison to the full calculation illustrates that in a broad range of $V_g$ it is enough to use a single-level model for describing the transport through the BDT molecule and suggests approaches based on a single-level model \cite{Baldea_ModelTrans_2015}. 
The structure in the stability diagram is a consequence of the strong coupling to the leads.
In the present  case the leads within channel 1 consist again of two nearly decoupled channels, the \quot{supporting channel} and the \quot{conducting channel}.
The real (imaginary) part $R^L$  ($\Gamma^L$)  of the retarded left lead hybridization function  $\Delta^{L,r}$ for these two channels are shown in \fref{fig:sdt}.
Only the \quot{conducting channel} has states above the Fermi level.
The real part $R^L$  shifts the levels in the central region.
{$R^L$ of the \quot{supporting channel} has anti-resonances at $\omega_1 = -0.5~$eV and $\omega_2 = -1.6~$eV.
The spike at $-1.3~$eV is due to the small coupling between the channels in the leads and its influence is negligible.
The imaginary part $\Gamma^L$ broadens all the levels and has maxima at the anti-resonances in $R^L$.
By the \quot{supporting effect} we mean that at every $V_b$, where 
the positive branch of an antiresonance  of $R_L$ 
hits the transport window of the opposite lead, 
the energy level in the central region will be shifted up to higher energies and therefore
one needs smaller $V_g$ to compensate this and the conductivity structure in the stability diagram bends down, see \fref{fig:SLmodel}.
The blue dashed lines show the case where the positive branches of the anti-resonances of $R_L$ hit the upper edge of the opposite lead,
\begin{eqnarray}
 \omega_1-\frac{V_b}{2} &= \Delta E_{empty} + \frac{V_b}{2} \\
 \omega_2-\frac{V_b}{2} &= \Delta E_{empty} + \frac{V_b}{2} \;,
\end{eqnarray}
and act as limit where the effect is maximal.
The same effect with opposite sign happens if a negative branch of an antiresonance of $R_L$ 
hits the transport window of the opposite lead, there the conductivity structure in the stability diagram bends up.
The red dashed line in \fref{fig:SLmodel} 
shows the case where the negative branch of the anti-resonances of $R_L$ hits the lower edge of the transport window, the Fermi energy, of the opposite lead,
\begin{equation}
 \omega_2-\frac{V_b}{2} = + \frac{V_b}{2}\;.
\end{equation}
There the structure in the stability diagram is maximally bent up.
This effect explains the appearance of more than one maximum in the current voltage characteristic even if there is only a single level in the central region.}

\begin{figure}[t]
\includegraphics[width=1.0\columnwidth,angle=0]{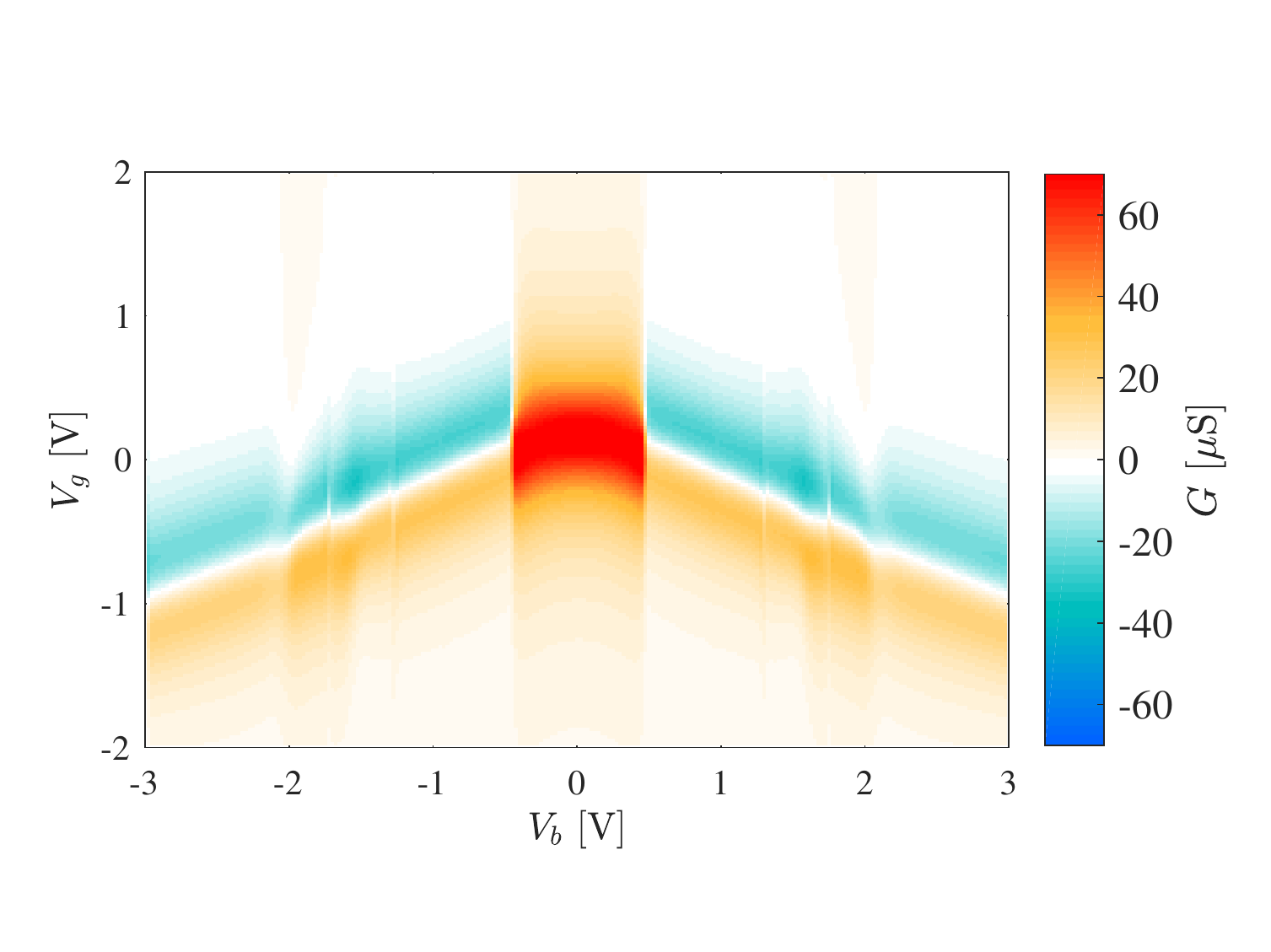}
\caption{Stability diagram based on the LDA data corresponding to the level at $-0.4~$eV. The supporting channel is manually blocked.}
\label{fig:ch1}
\end{figure}

Between $-\Delta E_{empty}< V_b < \Delta E_{empty}$ is the usual structure of a stability diagram, but smeared out due to the coupling of the leads.
Disabling the supporting channel produces the stability diagram in \fref{fig:ch1} where the structure can be explained just by the \quot{band edge effect}.

\subsection{Many-body effects in the Pt-BDT-Pt system}
\label{manybody}

After having discussed the properties of the single-particle part of the Hamiltonian, we now include many-body effects using the DFT+CPT method.
The interaction parameters $U_{ij}$ entering the model Hamiltonian in \eref{eq_hamiltonian} are determined by numerical integration of \eref{eq_U2} and listed  in \sref{parameters}.
We take a constant screening parameter $\eta = 1.5$ as proposed by Ryndyk et al. \cite{Ryndyk_BDT_2013}.
The values obtained in our calculation for the nearest neighbour hopping between the carbon atoms within benzene in \eref{eq:t} and the on-site Hubbard interaction in \eref{eq:U} are comparable to those reported in \cite{Bursill_PPPModel_1998}. 

In \tref{tab:eigenergies} the lowest many-body eigenenergies of the isolated central region are listed. 
The ground state ($10_{g}$) is in the $10$-particle sector and is a singlet state. The other low-lying states in the $10$-particle sector have singlet or triplet symmetry.
The lowest eigenenergies in the $9$-particle sector are doublet states.
\begin{table}[t]
	\begin{center}
\begin{tabular}{|c|c|c|c|c|c|}
\hline
Level & Energy [eV] & Spin [$\hbar$] & Level & Energy [eV] & Spin [$\hbar$] \\
 \hline
$10_g$ & -203.08 & 0 & &  &  \\
$10'$ & -200.81 & 1 & $9_g$ & -202.48 & 1/2 \\
$10''$ & -200.24 & 1 & $9'$ & -200.95 & 1/2 \\
$10'''$ & -200.11 & 0 & $9''$ & -199.46 & 1/2 \\
 \hline
\end{tabular}
	\caption{Many-body eigenenergies of the central region in case of Pt-BDT-Pt.
	\label{tab:eigenergies}
	}
	\end{center}
	\label{tab:eigenstates}
\end{table}

Including interaction in principle shifts levels and spectral weight away from the Fermi energy 
and produces additional peaks and therefore signatures in the current-voltage characteristic at higher voltages.
Therefore the benzene band gap increases compared to the LDA calculation \cite{Neaton_Benzene_2006} and can cause a reduction of the current.
As suggested by \cite{Pontes_AuBDTAu_2011, Sanvito_SIC_2007} the gap also widens in calculations based on  the self-interaction correction (SIC) in DFT.
In our transport system adding the full interaction  the HOMO level is shifted down to $-0.6~$eV, see DFT+CPT result in \fref{fig:transport}.
The peaks obtained by  this  calculation can be identified with excitations obtained by a true  many-body calculation for the effective Hubbard model.
There is a peak at $-0.6~$eV corresponding to the excitation from the $10_g$ to the $9_g$ state.
The next peak is at $-2.1~$eV and corresponds to the excitation from the $10_g$ to the $9'$ state.
This excitation needs more energy than we have taken into account  in our Wannier basis and are therefore negligible.

\begin{figure}[t]
\includegraphics[width=1.0\columnwidth,angle=0]{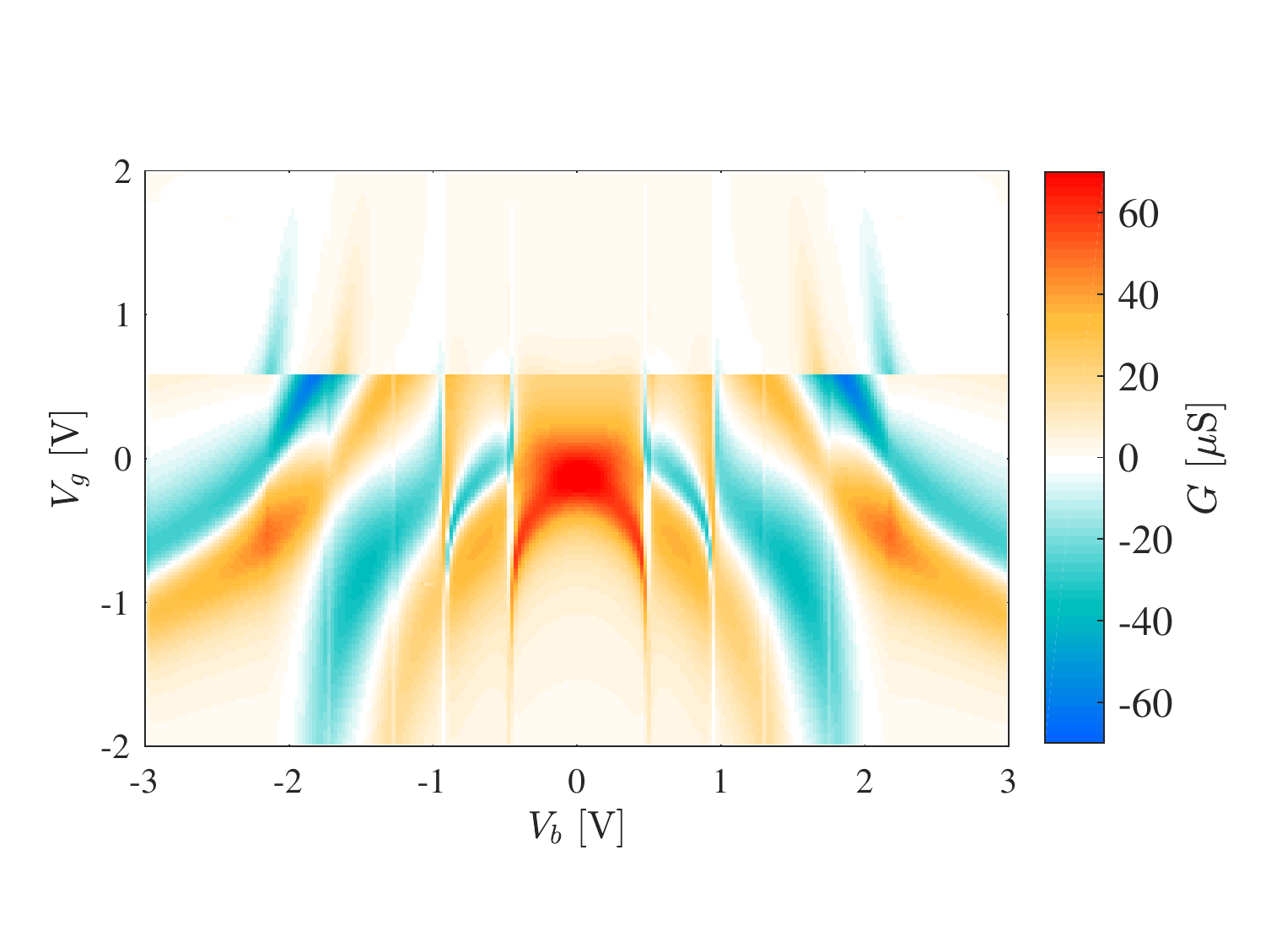}
\caption{Stability diagram based on  the DFT+CPT calculation.}
\label{fig:currentLDA+U}
\end{figure}

The stability diagram obtained by including electron-electron correlations within CPT does not change qualitatively for $V_g<0.6 ~$eV, see \fref{fig:currentLDA+U}.
The drastic change at $V_g = 0.6~$eV is an artefact of CPT.
\Eref{eq:Lehmann} is solved once for each $V_g$ independent of the applied $V_b$.
At $V_g = 0.6~$eV the HOMO level gets depleted and therefore the particle sector changes.
We expect that using more sophisticated non-equilibrium approaches like the ME+CPT calculation \cite{Nuss_MECPT_2015} the drastic change at $V_g = 0.6~$eV should disappear.
The conductance at $V_b=V_g=0$ of $0.86~G_0$ is higher than the result obtained with LDA even though the HOMO is shifted down in energy and has less spectral weight.
In the DFT+CPT calculation the spectral weight on the HOMO level is lower but distributed in a way that there is more weight in the orbitals near to the leads which causes the higher conductance.
\rumcomm{Thygesen et al. \cite{Thygesen_BDT_2011} have obtained the same trend at the BDT molecule connected with gold tips and studied with LDA and GW.}
In contrast to Ryndyk et al. \cite{Ryndyk_BDT_2013} the lead coupling effects overcome the influence of strong electron correlations in our transport system
because of stronger coupling between leads and BDT due to the geometry of the transport system.

\section{Conclusion}
\label{Conclusions}

We have performed first-principle calculations based on DFT, MLWFs, NEGF and CPT to study the molecular system consisting of benzenedithiolate connected to semi-infinite 
monoatomic metal electrodes.
DFT, within the plane-wave pseudopotential method, is used to calculate the electronic band structure of the transport system.
Transforming the Kohn-Sham eigenvalues and eigenfunctions to a real-space basis of Maximally Localized Wannier Functions allows to extract a tight-binding Hamiltonian to model the transport system.
Non-equilibrium Green's functions are  used in turn to calculate the charge transport through the BDT molecule.
In the case of  gold electrodes the HOMO level, which provides the dominant contribution to transport properties does not contribute to transport due to symmetry reasons and therefore the conductance is small.
Platinum electrodes, on the other hand, enable transport via the HOMO level.
Strong electron correlations are included on the BDT molecule using an extended Hubbard Hamiltonian.
Since the system is in the intermediate coupling regime
the spectral properties of the leads overcome 
the influence of many-body effects in this system.
We find that  \quot{band edge effects} and  \quot{supporting effects} are more relevant  for the structure of the stability diagram in this regime.

\section{Acknowledgements}

We  would  like  to  thank  M. Aichhorn for fruitful discussions. This work was partially supported by the Austrian Science Fund (FWF) within the SFB-ViCoM (F4103 and F4115) and P26508, and NAWI Graz.
The DFT calculations were partly done on high performance computing resources of the ZID at TU Graz.

\appendix

\section{Parameters of the model Hamiltonian}
\label{parameters}

The matrices in \eref{eq:VTLC}, \eref{eq:t} and \eref{eq:VCTR} are the single-particle parameters of the Pt-BDT-Pt system.
$\tilde{t}$ are the hopping parameters of the of the central region (C) and ${V}_{T_LC}$ and ${V}_{CT_R}$ the coupling matrices between the central region and the transition layers.
The row and column indices of $\tilde{t}$ correspond to the basis functions in \fref{fig:PtchIWan}.
\begin{equation}
 {V}_{T_LC} = \left( \begin{array}{cccccccc} -1.11 &   0.08 &   0.01  & -0.34  &  0.14  & -0.02  & -0.02  &  0.00  \\
   -0.85 &   0.11 &  0.00  &  0.08  & -0.05 &  -0.02  &  0.01  &  0.00 \end{array} \right)
   \label{eq:VTLC} 
\end{equation}
\begin{equation}
 \tilde{t} = \left( \begin{array}{cccccccc}
   -1.46  & -2.12  &  0.21  &  0.17  &  0.07  &       0  &  0.05 &  -0.03 \\
   -2.12  & -0.29  & -2.58  & -2.54  &  0.01  &  0.27  & -0.23 &   0.05 \\
    0.21  & -2.58  &  0.16  &  0.01  & -0.08  & -2.77  &  0.02 &   0.07 \\
    0.17  & -2.54  &  0.01  &  0.53  & -2.77  & -0.24  &  0.27 &        0 \\
    0.07  &  0.01  & -0.08  & -2.77  &  0.16  &  0.01  & -2.58 &   0.21 \\
         0  &  0.27  & -2.77  & -0.24  &  0.01  &  0.52  & -2.54 &   0.18 \\
    0.05  & -0.23  &  0.02  &  0.27  & -2.58  & -2.54  & -0.29 &  -2.12 \\
   -0.03  &  0.05  &  0.07  &       0  &  0.21  &  0.18  & -2.12 &  -1.46  \end{array} \right)
  \label{eq:t} 
\end{equation}
\begin{equation}
 {V}_{CT_R}^\dagger = \left( \begin{array}{cccccccc} 0.00  &  0.02  & -0.14  &  0.03 &  -0.01  &  0.34 &  -0.08  &  1.11 \\
   0.00  & -0.01   & 0.05   & 0.02  &  0.00  & -0.08  & -0.11  &  0.85 \end{array} \right)
   \label{eq:VCTR} 
\end{equation}
The interaction parameters $U_{ij}$ entering the model Hamiltonian determined by numerical integration of \eref{eq_U2} are
\begin{equation}
 U = \left( \begin{array}{cccccccc}  7.72  &  4.86  &  3.24  &  3.25  &  2.27   & 2.21  &  1.96 &   1.49 \\
    4.86 &  8.52   & 5.06  &  4.97  &  3.34   & 3.23   & 2.85  &  1.96  \\
    3.24 &   5.06  &  8.67 &   3.32 &   3.00  &  5.04  &  3.34 &   2.27 \\
    3.25 &   4.97  &  3.32 &   7.93 &   5.03  &  2.86  &  3.23 &   2.22 \\
    2.27 &   3.34  &  3.00 &   5.03 &   8.67  &  3.32  &  5.06 &   3.24 \\
    2.21 &   3.23  &  5.04 &   2.86 &   3.32  &  7.94  &  4.97 &   3.26 \\
    1.96 &   2.85  &  3.34 &   3.23 &   5.06  &  4.97  &  8.52 &   4.86 \\
    1.49 &   1.96  &  2.27 &   2.22 &   3.24  &  3.26  &  4.86 &   7.72 \\ \end{array} \right) \;.
    \label{eq:U}
\end{equation}
The relative integration error is within $5~$\%.

\section*{References}

\bibliography{paper}
\bibliographystyle{iopart-num}

\end{document}